\documentclass[twocolumn,aps,prl,showpacs, superscriptaddress, amsmath,amssymb]{revtex4}

\usepackage{graphicx}
\usepackage{dcolumn}
\usepackage{bm}
\usepackage{epstopdf}
\usepackage{color}
\usepackage{pdfpages}

\begin{document}

\title{Dynamics of ellipsoidal tracers in swimming algal suspensions}
\author{Ou Yang}
\affiliation{Department of Chemical Engineering and Materials
Science, University of Minnesota, Minneapolis, MN 55455}
\author{Yi Peng}
\affiliation{Department of Chemical Engineering and Materials
Science, University of Minnesota, Minneapolis, MN 55455}
\author{Zhengyang Liu}
\affiliation{Department of Chemical Engineering and Materials
Science, University of Minnesota, Minneapolis, MN 55455}
\author{Chao Tang}
\affiliation{Department of Chemical Engineering and Materials
Science, University of Minnesota, Minneapolis, MN 55455}
\author{Xinliang Xu}
\affiliation{Beijing Computational Science Research Center, Beijing 100193, China}
\author{Xiang Cheng}
\email{xcheng@umn.edu}
\affiliation{Department of Chemical Engineering and Materials
Science, University of Minnesota, Minneapolis, MN 55455}
\date{\today}

\begin{abstract}

Enhanced diffusion of passive tracers immersed in active fluids is a universal feature of active fluids and has been extensively studied in recent years. Similar to microrheology for equilibrium complex fluids, the unusual enhanced particle dynamics reveal intrinsic properties of active fluids. Nevertheless, previous studies have shown that the translational dynamics of spherical tracers are qualitatively similar, independent of whether active particles are pushers or pullers---the two fundamental classes of active fluids. Is it possible to distinguish pushers from pullers by simply imaging the dynamics of passive tracers? Here, we investigated the diffusion of isolated ellipsoids in algal {\it C. reinhardtii} suspensions---a model for puller-type active fluids. In combination with our previous results on pusher-type {\it E. coli} suspensions [Peng et al., Phys. Rev. Lett. {\bf 116}, 068303 (2016)], we showed that the dynamics of asymmetric tracers show a profound difference in pushers and pullers due to their rotational degree of freedom. Although the laboratory-frame translation and rotation of ellipsoids are enhanced in both pushers and pullers, similar to spherical tracers, the anisotropic diffusion in the body frame of ellipsoids shows opposite trends in the two classes of active fluids. An ellipsoid diffuses fastest along its major axis when immersed in pullers, whereas it diffuses slowest along the major axis in pushers. This striking difference can be qualitatively explained using a simple hydrodynamic model. In addition, our study on algal suspensions reveals that the influence of the near-field advection of algal swimming flows on the translation and rotation of ellipsoids shows different ranges and strengths. Our work provides not only new insights into universal organizing principles of active fluids, but also a convenient tool for detecting the class of active particles.                

\end{abstract}

\maketitle

\section{I. Introduction}

Active fluids are a novel class of nonequilibrium soft materials, which are composed of a large number of self-propelled particles suspended in simple fluids \cite{Ramaswamy10,Koch11,Marchetti13,Schwarz-Linek16,Bechinger16}. The self-propelled particles can convert ambient or internal free energy into persistent motions with a direction depending on the local configuration of particles and interparticle interactions. Suspensions of swimming microorganisms and synthetic colloidal swimmers are the most widely studied active fluids in experiments and frequently serve as models for theoretical investigations \cite{Schwarz-Linek16,Ebbens10}. Joint efforts of experiments, simulations, and theories have shown that active fluids exhibit surprising behaviors such as giant number fluctuations \cite{Narayan07,Palacci13,Zhang13}, ordered phases with collective particle motions \cite{Marchetti13,Sokolov12,Wensink12}, and abnormal rheology \cite{Sokolov09,Rafai10,Gachelin13,Lopez15}, unknown to conventional equilibrium complex fluids.  Among these features, the enhanced diffusion of passive tracers in active fluids has attracted probably the most extensive and sustained interests in recent years. 

Wu and Libchaber first showed that a spherical tracer immersed in suspensions of swimming {\it Escherichia coli} exhibits a superdiffusive motion at short times and an enhanced diffusion at long times \cite{Wu00}. Such an enhanced diffusion can be orders of magnitude stronger than the tracer's intrinsic Brownian motion at high bacterial concentrations. Following their pioneering work, the enhanced diffusion of passive spherical tracers has been reported and systematically studied in different active fluids including suspensions of swimming microorganisms such as prokaryotic cells {\it E. coli} \cite{Kim04,Chen07,Mino11,Wilson11,Mino13,Jepson13,Patteson16}, {\it Bacillus subtilis} \cite{Valeriani11} and {\it Pseudomonas
sp.} \cite{Vaccari15} and eukaryotic cell {\it Chlamydomonas reinhardtii} \cite{Leptos09,Kurtuldu11} as well as synthetic colloidal microswimmers \cite{Mino11}. The motion of passive tracers in active fluids is induced by the fluid flow of microswimmers in the far field and the direct steric interaction between tracers and microswimmers in the near field \cite{Mino13,Underhill08,Ishikawa10,Lin11,Pushkin13_1,Pushkin13_2,Morozov14,Mathijssen15,Yeo16}. Thus, similar to microrheology in equilibrium systems \cite{Squires10}, the study of the dynamics of passive tracers can reveal the intrinsic properties of active fluids. In addition, understanding the motion of passive tracers in suspensions of swimming microorganisms is also of biological relevance \cite{Wu00,Morozov14}. The enhanced diffusion of passive particles such as nutrient granules, dead bacterial bodies, and liquid droplets of macromolecules boosts fluid mixing at microscopic scales \cite{Kurtuldu11,Katija12}, which helps to maintain an active ecological balance and promotes intercellular signaling and metabolite transport. 

Although the dynamics of {\it spherical} passive tracers have been extensively studied \cite{Wu00,Kim04,Chen07,Mino11,Wilson11,Mino13,Jepson13,Kasyap14,Patteson16,Valeriani11,Vaccari15,Leptos09,Kurtuldu11,Underhill08,Ishikawa10,Lin11,Pushkin13_1,Pushkin13_2,Morozov14,Mathijssen15}, the motion of {\it asymmetric} tracers that possess degrees of freedom beyond simple translation has not been investigated until very recently. Peng {\it et al.} studied the dynamics of isolated ellipsoids in {\it E. coli} suspensions and showed that both the translational and rotational diffusion of ellipsoids are enhanced with increasing bacterial concentrations \cite{Peng16}. More importantly, they found an abnormal anisotropic diffusion in the body frame of ellipsoids, where an ellipsoid diffuses fastest along its minor axis at high bacterial concentrations. Such an abnormal anisotropic diffusion leads to a negative translation-rotation coupling in the laboratory frame that is strictly forbidden for Brownian particles. Based on a simple mean-field hydrodynamic calculation, they argued that the unusual anisotropic diffusion is a result of the universal straining flow created by pusher-type active particles with {\it E. coli} as one specific example. Moreover, the calculation predicts that asymmetric tracers immersed in puller-type active fluids should show an opposite trend, where the diffusion of ellipsoids along the major axis should be more strongly enhanced. However, such a prediction has not been experimentally verified.

\begin{figure}
\begin{center}
\includegraphics[width=3.2in]{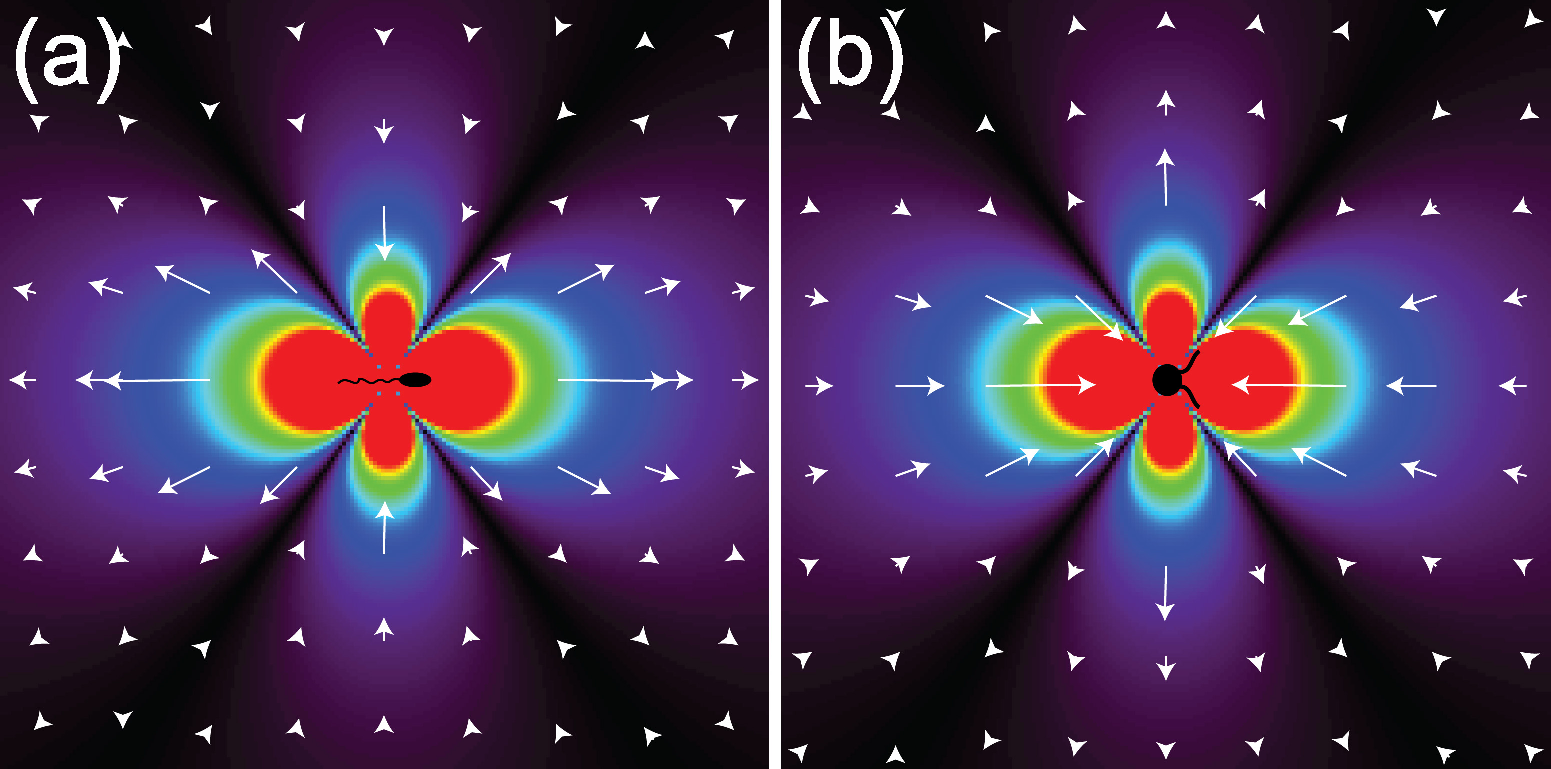}
\end{center}
\caption[Pusher and Puller]{Schematics showing the flow patterns induced by a single pusher (a) and a single puller (b). The color indicates the magnitude of the flow, which decays from the origin following a $r^{-2}$ scaling. The arrows indicates the direction of the flow. {\it E. coli} is a typical pusher-type microswimmer \cite{Peng16}, whereas {\it C. reinhardtii} we study here is a typical puller-type microswimmer.} \label{figure1}
\end{figure}

Depending on the forces exerted on the surrounding fluid, isolated self-propelled microswimmers can be generally categorized in two classes, i.e., ``pushers'' and ``pullers'' \cite{Koch11,Marchetti13}. In the first order of a multipole expansion, a pusher creates a tensile dipole flow in the far field, where the fluid is pushed out parallel to the swimming direction of the swimmer and is pulled in at its midpoint perpendicular to the swimming direction (Fig.~\ref{figure1}(a)). In contrast, a puller creates a contractile dipole flow, where the fluid is pulled in parallel to the swimming direction and is pushed out at its midpoint (Fig.~\ref{figure1}(b)). Most bacteria including {\it E. coli} and {\it B. subtilis} propel by rotating long thin flagella, which push fluid backward and, therefore, are pushers \cite{Drescher11}. Algae such as {\it C. reinhardtii} propel by beating two anterior flagella that pull in fluid in the front and are well-known examples of pullers \cite{Drescher10,Guasto10}. Although the two classes of active particles lead to profound differences in the behaviors of active fluids, such as their rheological response under shear \cite{Hatwalne04,Cates08,Giomi10} and their stability in ordered swarming phases \cite{Ramaswamy10,Underhill08,Saintillan08,Baskaran09}, the influences of pushers and pullers on the enhanced diffusion of spherical tracers are identical in the dilute limit \cite{Vaccari15,Kurtuldu11,Lin11,Pushkin13_1}. In both cases, the enhanced diffusion of passive tracers increases linearly with the active flux, defined as the product of the number density and the speed of microswimmers \cite{Mino11,Mino13,Jepson13,Leptos09,Ishikawa10,Underhill08,Lin11,Pushkin13_1}.  

In contrast, the calculation by Peng {\it et al.} showed that when the rotational degree of freedom is considered, the swimming mechanism of microswimmers directly affects the coupling between the translation and rotation of asymmetric passive tracers. Therefore, it is possible to distinguish the swimming mechanism of active particles by simply imaging the diffusion of asymmetric tracers immersed in active fluids. The method should be easier and faster to implement when compared with other commonly used procedures for probing the swimming mechanism of microswimmers such as imaging the periodic time-irreversible stroke of flagella or cilia of microorganisms \cite{Turner00,Polin09} or mapping the far-field flow field through particle imaging velocimetry (PIV) \cite{Drescher11,Drescher10,Guasto10}. For instance, the width of a single flagellum of {\it E. coli} is $\sim 20$ nm, which cannot be observed using optical microscopy without special treatments \cite{Turner00}. It is hard, if not impossible, to image the concentration gradient of various chemical species around the body of self-propelled catalytic Janus particles, which can be either pushers or pullers depending on the area fraction of the active catalytic region on the surface of the particles \cite{Sharifi-Mood15}. Although PIV can resolve the strength of multipole flow fields to higher orders, it requires a large number of measurements to reduce statistical noise and demands manipulation of the position and orientation of individual microswimmers for image analysis \cite{Drescher11,Drescher10,Guasto10}. Hence, the study of the dynamics of asymmetric particles in active fluids will provide not only new insights into the universal features of active fluids, but also a convenient tool for characterizing the fundamental properties of active particles.        
      
Although the dynamics of spherical tracers have been studied in both pushers and pullers  \cite{Wu00,Kim04,Chen07,Mino11,Wilson11,Mino13,Jepson13,Kasyap14,Patteson16,Valeriani11,Vaccari15,Leptos09,Kurtuldu11,Underhill08,Ishikawa10,Lin11,Pushkin13_1,Pushkin13_2,Morozov14,Mathijssen15}, the dynamics of asymmetric tracers have only been investigated in pusher-type active fluids so far \cite{Peng16}. Here, we reported our study on the dynamics of isolated ellipsoids in a premier puller-type active fluid---suspensions of algae {\it C. reinhardtii}. We studied the rich dynamics of ellipsoids in algal suspensions and drew a detailed comparison between the dynamics of ellipsoids and spherical tracers in both puller-type and pusher-type active fluids. Particularly, we investigated the anisotropic diffusion of ellipsoids in the body frame and verified the prediction of the simple hydrodynamic model. Moreover, thanks to the large size of {\it C. reinhardtii}, the algal system allows us to probe the detail of the near-field translational and rotational advection of ellipsoidal tracers under the influence of single alga that cannot be easily achieved with {\it E. coli} \cite{Leptos09}. Through this study, we show that the influence of an algal swimming flow on the rotation of ellipsoids is weaker and has a shorter range, when compared with the influence of the same flow on the translation of ellipsoids. We provide a quantitative estimate of the range of influences on different degrees of freedom.        

Our paper is organized as follows. We introduce materials and experimental methods in Section II and show results and discussions in Section III. Specifically, the laboratory-frame dynamics and the body-frame dynamics of ellipsoids are presented in two separate subsections of Section III. The anisotropic diffusion of ellipsoids---the key feature that distinguishes between pusher-type and puller-type active fluids---is then discussed in subsection III C. Conclusions are summarized in Section IV.   

\begin{figure}
\begin{center}
\includegraphics[width=3.2in]{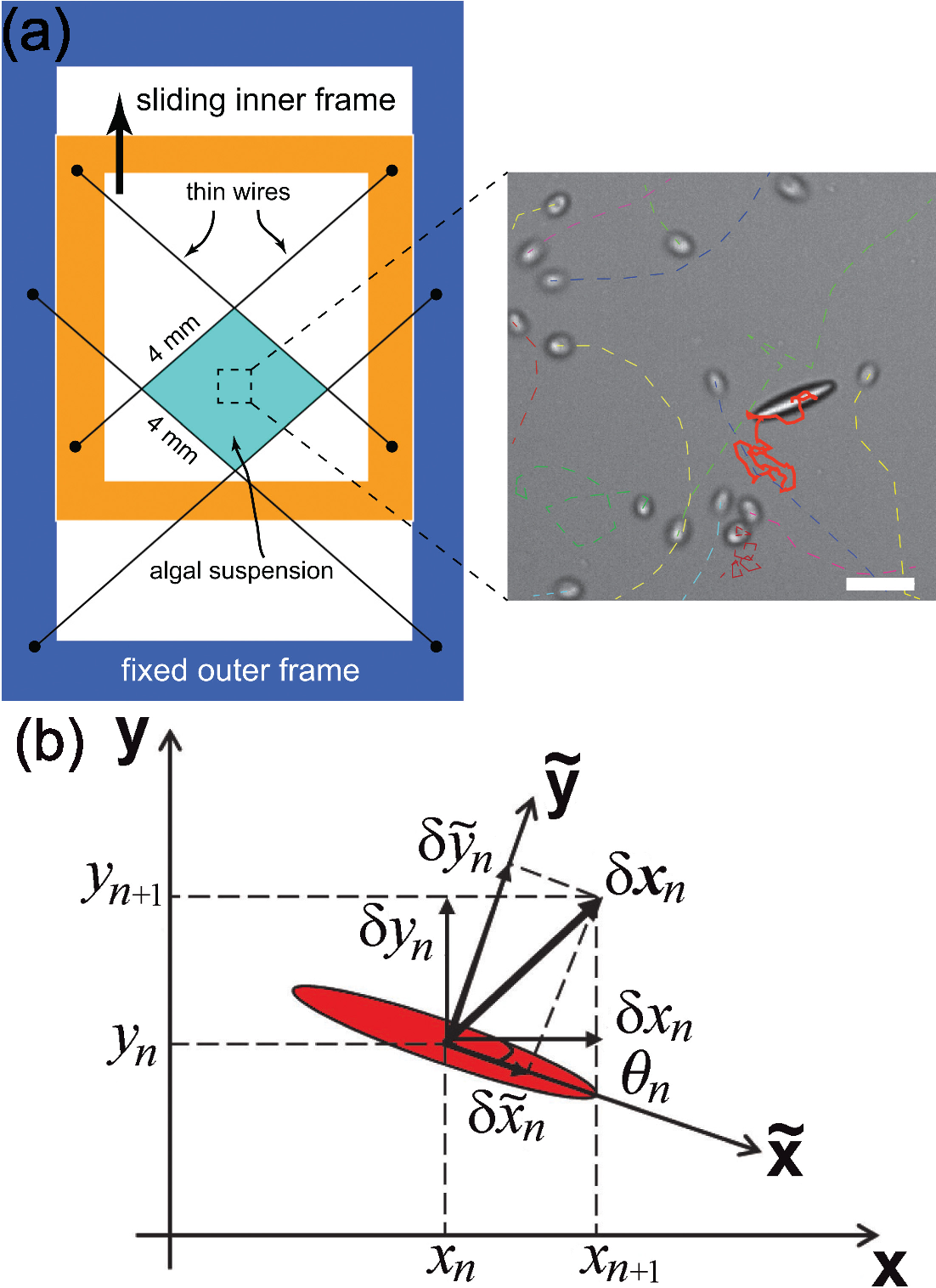}
\end{center}
\caption[Setup and methods]{Experimental setup and method. (a) The left panels show a schematic of the adjustable wire-frame device. The right panel shows a zoomed view of a region of $130 \times 130$ $\mu$m$^2$ in the center of the free-standing film. The solid red line indicates the trajectory of the ellipsoid over a time interval of 10 s. The dashed lines indicate the trajectories of algae. Scale bar: 20 $\mu$m. (b) Reference frame transformation. The displacement vector, $\delta \mathbf{x}_n$, of an ellipsoid in a small time interval can be decomposed into two orthogonal components ($\delta x_n$, $\delta y_n$) in the laboratory frame based on the $x$-$y$ axis or ($\delta \tilde{x}_n$, $\delta \tilde{y}_n$) in the body frame based on the $\tilde{x}$-$\tilde{y}$ axis. The two coordinates are connected through a 2D rotational matrix specified by the orientation of the ellipsoid in the laboratory frame, $\theta_n$.} \label{figure2}
\end{figure}

\section{II. Experiment}

\subsection{A. Materials}
We used wild-type {\it C. reinhardtii} (CC125+) as our active particles. {\it C. reinhardtii} is a unicellular green alga with a spherical body of diameter $d \approx$ 7--10 $\mu$m. It has two anterior flagella about 10--12 $\mu$m in length. The flagella beat at $\sim$ 50 Hz and propel the cell at a mean speed of 100--200 $\mu$m/s \cite{Harris89}. We cultured algae in minimal media (M1) on a light cycle of 14 h bright and 10 h dark. The procedure helps to increase the uniformity of the size and the speed of algae \cite{Kurtuldu11}. To vary the concentration of algal suspensions, we centrifuged suspensions at 1800 rpm for 1.5 min and then resuspended the concentrated suspensions in M1 media to desired concentrations.  

We used polystyrene (PS) ellipsoids as our passive asymmetric tracers. The ellipsoids were obtained by stretching monodispersed micron-sized PS spheres at 150 $^\circ$C above the glass transition temperature of PS \cite{Ho93}.  The lengths of the semi-principal axes of the ellipsoids were fixed at $a = b = 2.8 \pm 0.2$ $\mu$m and $c = 14.2 \pm 0.5$ $\mu$m, the same as the ellipsoids used in the previous study on pusher-type {\it E. coli} suspensions \cite{Peng16}. The aspect ratio of ellipsoids is $p\equiv c/a = 5.1$.  A small number of PS ellipsoids were mixed into algal suspensions with the volume fraction of ellipsoids below $0.05\%$. The concentration is so low that the hydrodynamic coupling between ellipsoids can be safely neglected in our experiments.

\subsection{B. Methods}
A drop of an algal suspension containing PS ellipsoids was first deposited onto a small gap suspended by four thin wires made of human hairs. By enlarging the distance between the four wires, we stretched the suspension into a free-standing $4 \times 4$ mm$^2$ liquid film with a thickness $\sim 20$ $\mu$m (Fig.~\ref{figure2}(a)). The construction of the adjustable wire-frame device is similar to those used in previous experiments on enhanced diffusions of passive tracers \cite{Sokolov12,Guasto10,Kurtuldu11,Peng16}. To stabilize the thin liquid film, a trace amount of surfactant (Tween 20, $0.03$ vol$\%$) was added into the algal suspensions. We quantify the concentration of algae in the thin film by measuring the area fraction of algae $\phi=N\pi\langle d\rangle^2/4A$, where $N$ is the number of algae in the field of view, $\langle d\rangle = 8$ $\mu$m is the mean diameter of algae, and $A$ is the total area of the field of view. Note that, differently from $E. coli$ suspensions where high bacterial concentrations can be achieved \cite{Peng16}, the maximal $\phi$ of algal suspensions is limited in the dilute regime \cite{Leptos09}. When $\phi$ is above $4\%$, {\it C. reinhardtii} start to shed flagella and stop swimming in our experiments.

We recorded the motions of algae and ellipsoids at a frame rate of 10 frames/second using a Nikon Ti-E inverted microscope with $\times20$ 0.5 NA (numerical aperture) objective (Fig.~\ref{figure2}(a) and Supplemental Movies 1 and 2 \cite{Yang16}). To eliminate the phototaxis of algae, a long-pass filter with a cut-on wavelength of 620 nm was used for illumination. We extracted the position and orientation of ellipsoids using a custom particle tracking algorithm. The laboratory-frame trajectories of the center of mass of ellipsoids, $\mathbf{x}(t)$, and the orientation of ellipsoids, $\theta(t)$, can then be obtained. Here, $\theta$ is defined as the angle of the major axis of ellipsoids with respect to the $x$ axis, arbitrary selected and fixed in the laboratory frame. Due to the centrosymmetry of ellipsoids, $\theta$ is limited between $-\pi/2$ and $\pi/2$.

In order to probe the anisotropic diffusion of ellipsoids, we transformed the motion of ellipsoids from the laboratory frame into the body fame (Fig.~\ref{figure2}(b)). The body frame is a special frame of reference that rotates along with ellipsoids. Particle displacements in the body frame were obtained through rotation of particle displacements in the laboratory frame (Fig.~\ref{figure2}(b)) \cite{Han06,Mukhija07,Peng16}. Specifically, the displacement of an ellipsoid in the laboratory frame, $\delta \mathbf{x}(t_n) =\mathbf{x}(t_{n+1})-\mathbf{x}(t_n)$, in a small time interval $\delta t = t_{n+1}-t_n$ was transformed into its displacement in the body frame, $\mathbf{\delta \tilde{x}}(t_n)$, via $\mathbf{\delta \tilde{x}}(t_n)=R(t_n)\delta \mathbf{x}(t_n)$, where $R(t_n)$ is a 2D rotation matrix with $R(t_n)=\left(\begin{matrix}
\cos\theta_n & \sin\theta_n \\
-\sin\theta_n & \cos\theta_n
\end{matrix} \right)$. Here, $\theta_n=[\theta(t_n)+\theta(t_{n+1})]/2$ is the average orientation of the ellipsoid in $\delta t$. Since the rotation of ellipsoids in $\delta t$ is small, choosing either $\theta_n = \theta(t_n)$ or $\theta_n=\theta(t_{n+1})$ does not change the final result \cite{Han06,Peng16}.  Last, we constructed the total body-frame displacement by summing displacements in each small time step, $\Delta\mathbf{\tilde{x}}(t)=\sum_{n=0}^{k}\delta\mathbf{\tilde{x}}(t_n)$, where $t_k = t_0 + t$. $\Delta\mathbf{\tilde{x}}(t)$ has two orthogonal components, $\Delta\tilde{x}$ and $\Delta\tilde{y}$, indicating the displacements of ellipsoids along the major and minor axes, respectively.

\section{III. Results and discussions}

\begin{figure*}
\begin{center}
\includegraphics[width=7in]{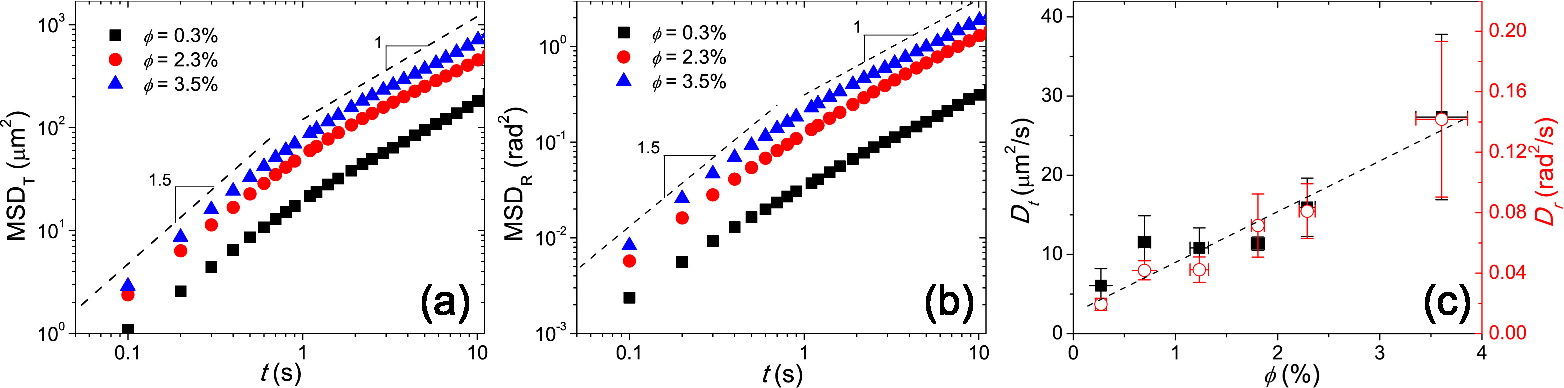}
\end{center}
\caption[Dynamics of ellipsoids in the laboratory frame]{Dynamics of ellipsoids in the laboratory frame. (a) Mean-square translational displacements of ellipsoids (MSD$_\text{T}$) at different algal concentrations $\phi$. (b) Mean-square rotational displacements of ellipsoids (MSD$_\text{R}$) at the same $\phi$. The dashed lines indicate the slope of the data. (c) Effective translational diffusivity, $D_T$ (the left axis), and effective rotational diffusivity, $D_R$ (the right axis), as a function of $\phi$. $D_T$ and $D_R$ were extracted from the long-time diffusions of the corresponding MSDs. The dashed line indicates a linear relation between $D_{T,R}$ and $\phi$.} \label{figure3}
\end{figure*}

\subsection{A. Dynamics in the laboratory frame}
We first investigated the dynamics of ellipsoids in the laboratory frame. Figure \ref{figure3}(a) and (b) show translational mean-squared displacements (MSD$_\text{T}$), $\langle\lbrack\Delta \mathbf{x}(t)\rbrack^2\rangle =\langle\lbrack \mathbf{x}(t+t_0)-\mathbf{x}(t_0)\rbrack^2\rangle$, and rotational mean-squared displacements (MSD$_\text{R}$), $\langle\lbrack \theta(t)\rbrack^2\rangle = \langle\lbrack \theta(t+t_0)-\theta(t_0)\rbrack^2\rangle$, respectively. In both cases, the average is taken over the initial time $t_0$. Similar to the dynamics of ellipsoids in pusher-type active fluids \cite{Peng16}, ellipsoids in algal suspensions show superdiffusive motions at short times and diffusive motions at long times. The effective translational and rotational diffusivity, $D_t$ and $D_r$, can be extracted by fitting the long-time diffusions with $\langle\lbrack\Delta \mathbf{x}(t)\rbrack^2\rangle = 4D_Tt$ and $\langle\lbrack \theta(t)\rbrack^2\rangle = 2D_Rt$. We found that both $D_T$ and $D_R$ increase linearly with algal concentrations (Fig.~\ref{figure3}(c)). It is worth noting that the intrinsic Brownian translation and rotation of the large ellipsoids used in our experiments are orders of magnitude smaller, even when compared with the diffusivity of ellipsoids in algal suspensions of the lowest concentration we studied. Indeed, we barely observed any diffusive motions of ellipsoids without algae. The weak diffusion led to large experimental uncertainties when we attempted to measure the bare diffusivity of ellipsoids in the liquid films. Instead, we estimated the Brownian diffusivity from the Stokes-Einstein relation and the anisotropic drag coefficients of ellipsoids in three dimensions \cite{Han09}, which gives $D_{T0} = 0.038$ $\mu$m$^2$/s and $D_{R0} = 3.1 \times 10^{-4}$ rad$^2$/s. 

The linear relationship between diffusivity and algal concentrations has also been found for enhanced diffusions of spherical tracers in both pushers and pullers including {\it E. coli} and algal suspensions \cite{Mino11,Mino13,Jepson13,Leptos09,Ishikawa10,Underhill08,Lin11,Pushkin13_1}. However, the result is different from a measurement on the enhanced diffusion of spherical tracers in quasi-two-dimensional algal suspensions, where the translational diffusivity increases nonlinearly with $\phi$ following $D_T\sim \phi^{3/2}$ \cite{Kurtuldu11}. The difference is unlikely due to the additional rotational degree freedom and probably arises from the three-dimensional (3D) nature of our experiments, where we used thicker liquid films and smaller algae. It is known that $D_T\sim\phi$ for spherical tracers in 3D algal suspensions \cite{Leptos09}. Moreover, $D_T$ is also $\sim \phi$ for spherical tracers in dilute {\it E. coli} suspensions confined in liquid films with a thickness of a couple of bacterial body lengths \cite{Wu00}. Last, for ellipsoids in {\it E. coli} suspensions, both $D_T$ and $D_R$ increase linearly with bacterial concentrations in the dilute limit \cite{Peng16}, qualitatively similar to our results in algal suspensions. 

\begin{figure}
\begin{center}
\includegraphics[width=3.2in]{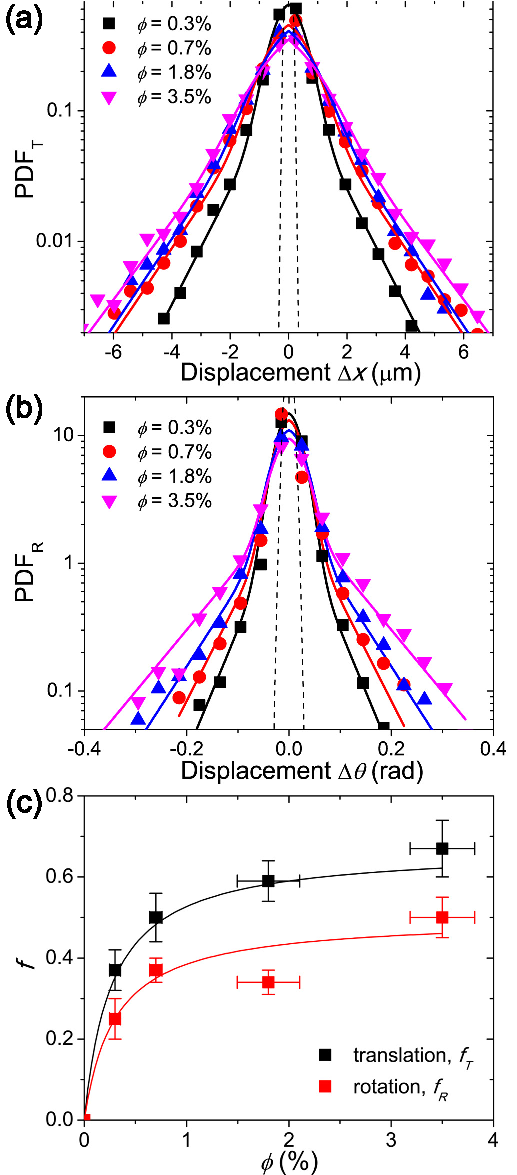}
\end{center}
\caption[PDFs of ellipsoids in the laboratory frame]{Probability distribution functions of laboratory-frame translational displacements (PDF$_\text{T}$) (a) and rotational displacements (PDF$_\text{R}$) (b) in a time interval $\Delta t = 0.1$ s at different algal concentrations $\phi$. The dashed lines are the PDFs of Brownian ellipsoids with bare translational and rotational diffusivity given in the text. The solid lines are fits using Eq.~\ref{equation1}. (c) The weighting factor, $f$, as a function of $\phi$ for both PDF$_\text{T}$ and PDF$_\text{R}$. $f$ indicates the relative contributions of diffusion and advection. The solid lines are visual guides.} \label{figure4}
\end{figure}

We also studied the probability distribution of translational and rotational displacements within a small time interval $\Delta t = 0.1$ s. Figures \ref{figure4}(a) and (b) show respectively the translational probability distribution function (PDF$_\text{T}$) and the rotational probability distribution function (PDF$_\text{R}$) at different algal concentrations. Both PDF$_\text{T}$ and PDF$_\text{R}$ show Gaussian cores at small displacements and long exponential tails at large displacements. The width of the distributions broadens significantly with increasing algal concentrations. The exponential tails of the large displacements are induced by the advection of the fluid flow of a single algae swimming close to ellipsoids, whereas the Gaussian cores indicate an effective diffusion induced by the average fluid flow of numerous algae further away from ellipsoids \cite{Leptos09}. It should be emphasized that although it is unambiguous that the large displacements of ellipsoids result from close encounters between individual algae and ellipsoids, the effective diffusion of the small displacements is also a direct consequence of algal flows instead of the thermal fluctuation of the surrounding media. In fact, the PDFs derived from the bare diffusivity $D_{T0}$ and $D_{R0}$ are significantly narrower than the PDFs of the lowest-concentration algal suspensions (Fig.~\ref{figure4}(a) and (b)). 

The relative contributions of the near-field advection and the far-field diffusion can be further quantified using a formula originally proposed for spherical tracers \cite{Leptos09}, 
\begin{equation}
\text{PDF}(\Delta \lambda) = \frac{(1-f)}{2\pi\delta_g^2}e^{-(\Delta \lambda)^2/2\delta_g^2} + \frac{f}{2\delta_e}e^{-|\Delta \lambda|/\delta_e},
\label{equation1}
\end{equation}  
where $\Delta \lambda \equiv \Delta x$ for translation and $\Delta \theta$ for rotation, $\delta_g$ indicates the standard deviation of the Gaussian core and $\delta_e$ indicates the decay length of the exponential tails. The weighting factor, $f(\phi) \in [0,1]$, quantifies the relative contributions of the Gaussian and the exponential terms. $f = 0$ gives a pure Gaussian distribution and $f = 1$ gives an exponential (Laplace) distribution. $f(\phi)$ extracted from both PDF$_\text{T}$ and PDF$_\text{R}$ increases with the algal concentration $\phi$ (Fig.~\ref{figure4}(c)), indicating an increasing influence of the near-field advection. Although $f$ from PDF$_\text{T}$ and PDF$_\text{R}$ follow a qualitatively similar trend, the rotational $f$, $f_r$, is consistently smaller than the translational $f$, $f_T$. At high $\phi$, $f_T$ gradually grows to a value around 0.67, whereas $f_R$ approaches 0.5. Based on the study of spherical tracers, Leptos {\it et al.} proposed a heuristic physical picture, where they suggested that there exists a sphere of influence surrounding each swimming alga \cite{Leptos09}. Tracers within the sphere are dominantly influenced by the advection induced by the fluid flow of the alga, whereas outside the sphere tracer dynamics are affected by the fluid flows of multiple algae, which on average lead to diffusive motions. The radius of the sphere is estimated as $r \sim \langle d\rangle (f/\phi)^{1/3}$ based on dimensional analysis, where $\langle d\rangle \approx 8$ $\mu$m is the average diameter of algae. Our results suggest that the sphere of influence for translational advection is systematically larger than that for rotational advection. The size ratio between the two spheres of influence at high $\phi$ is $r_T/r_R = (f_T/f_R)^{1/3} = 1.10 \pm 0.06$.                            

The difference between the sizes of the sphere of influence for the two degrees of freedom can be understood from the nature of the swimmer's velocity field. Without external forces, the fluid velocity around a single swimming microorganism, $u(r)$, follows a dipole form \cite{footnote1}, which decays as $u \sim r^{-2}$ in 3D \cite{Ramaswamy10,Koch11}. The velocity field of the dipole flow determines the translational motion of ellipsoids. On the other hand, the rotational motion of ellipsoids is dictated by the rate of strain and the vorticity of the dipole flow, which decays faster following $\Omega \sim r^{-3}$, where $\Omega$ is the vorticity \cite{Kim91}. Hence, the influence of the rotational advection imparted by swimming algae has a shorter range, leading to a smaller sphere of influence. Quantitatively, the oscillatory flow field around a single alga can be approximated as $u(r)\exp(i\omega t)$, where $\omega \approx 2\pi \times 50$ rad/s is the beating frequency of the flagella of algae \cite{Leptos09,Guasto10}. The radius of the sphere of influence in translation can be estimated as the location where the translational displacement due to the advection in a half cycle, $2u(r_T)/\omega$, is comparable with the diffusive displacement in the same time interval, $(2D_Tt)^{1/2} = (2\pi D_T/\omega)^{1/2}$ \cite{Leptos09}. Note that $t = \pi/\omega$. The argument leads to $u(r_T) \sim (\pi D_T \omega/2)^{1/2}$. Similarly, the radius of the sphere of influence in rotation can be estimated as $\Omega(r_R) \sim (\pi D_R \omega/2)^{1/2}$. From the dipole flow of swimming microorganisms, we have $u(r) \sim k/r^2$ and $\Omega(r) \sim \frac{1}{2}|\nabla \times \mathbf{u}| \sim k/r^3$, where $k$ is the dipole strength depending on the specific swimming mechanism of microorganisms. By using the relation $u(r_T)/\Omega(r_R) = (D_T/D_R)^{1/2}$, we reach 
\begin{equation}
\frac{r_T}{r_R} = \left(\frac{r_T^2 D_R}{D_T}\right)^{1/6}.
\label{equation2}
\end{equation}  
The radius of the sphere of influence in translation, $r_T$, depends on the dipole strength of algae, which can be estimated as $k = U_0l^2 \approx 4.9 \times 10^4$ $\mu$m$^3$/s \cite{footnote2}. Here, $U_0 \approx 150$ $\mu$m/s is the characteristic speed of algae from our direct measurements. A similar value has also been reported in Ref. \cite{Leptos11}. $l$ is the length of the force dipole. $l = \langle d \rangle + l_{f} \approx 18$ $\mu$m, where $\langle d \rangle$ is the average length of algae and $l_f \approx 10$ $\mu$m is the length of algal flagella. Using the relation $k/r_T^2 = (\pi D_T \omega/2)^{1/2}$ and $D_T \approx 27$ $\mu$m$^2$/s at high $\phi$ from our measurements (Fig.~\ref{figure3}(c)), we have $r_T \approx 20.5$ $\mu$m \cite{footnote3}.  Inserting $r_T$, $D_T$ and $D_R \approx 0.14$ rad$^2$/s from our measurements at high $\phi$ into Eq.~\ref{equation2}, we have $r_T/r_R \approx 1.14$, consistent with our estimate from $f$. Finally, as an interesting comparison, we can also calculate the ratio of length scales, $r_T'/r_R'$. $r_T'$ and $r_R'$ are the lengths where the advection in a half cycle is comparable with the size of tracers. Previous work has shown that $r_T'$ is the same order of magnitude as $r_T$ \cite{Leptos09}. For translation, $r_T'$ can be obtained from $2u(r_T')/\omega \approx \langle a \rangle$, where we use the average length of the three semi-axes of ellipsoids, $\langle a \rangle = (a+b+c)/3 = 6.6$ $\mu$m, as the characteristic size of ellipsoids. Similarly, we also have $2\Omega(r_R)/\omega \approx \pi/2$ for rotation, where we set $\pi/2$ as a typical angle of rotation. Combining these two relations, we have $r_T'/r_R' = \lbrack \pi r_T'/(2\langle a \rangle)\rbrack^{1/3} = 1.17$, quantitatively similar to $r_T/r_R$ we obtained above. Note that we use $r_T' \approx 6.8$ $\mu$m here, which is obtained from the relation $2k/(r_T'\omega) = \langle a \rangle$.                    

The laboratory-frame dynamics of ellipsoids show that the translational and rotational motions of ellipsoids arise from the same origin, i.e. the dipole flow of swimming algae. In comparison with previous studies on passive tracers in active fluids, we found that the short-time superdiffusion and the long-time diffusion of ellipsoids in algal suspensions are qualitatively similar to the dynamics of ellipsoids in pusher-type bacterial suspensions. Moreover, the linear relation between the enhanced diffusivity and algal concentrations is also quantitatively the same as the enhanced translational diffusion of spherical tracers in both pushers and pullers. Thus, if treated independently in the laboratory frame, the translational and rotational degrees of freedom of ellipsoids show identical behaviors regardless of the swimming mechanism of active particles. However, the nature of dipole flow dictates a coupling between the two degrees of freedom as we shall show next.  

\begin{figure*}
\begin{center}
\includegraphics[width=7in]{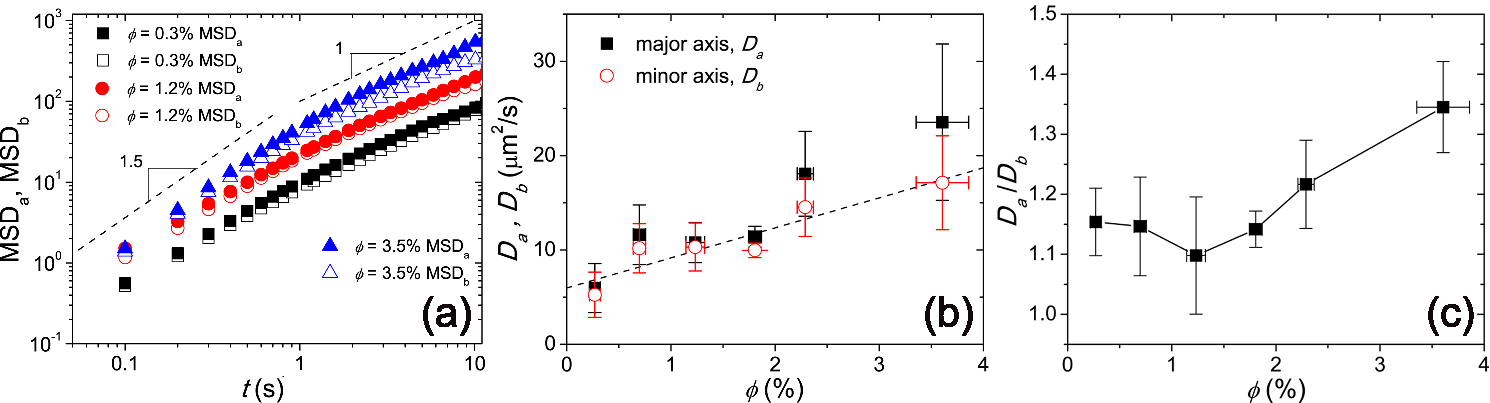}
\end{center}
\caption[Dynamics of ellipsoids in the body frame]{Dynamics of ellipsoids in the body frame. (a) Mean-square translational displacements of ellipsoids along the major axis (MSD$_\text{a}$) and along the minor axis (MSD$_\text{b}$) at different algal concentrations $\phi$. (b) Effective diffusivity along the major axis, $D_a$, and along the minor axis, $D_b$, as a function of $\phi$. $D_a$ and $D_b$ were extracted from the long-time diffusions of MSD$_\text{a}$ and MSD$_\text{b}$. The dashed line is a linear fit of $D_b$ for guiding the eye. (c) Anisotropic diffusion of ellipsoids, $D_a/D_b$, as a function of $\phi$.} \label{figure5}
\end{figure*}   

\subsection{B. Dynamics in the body frame}
The body-frame dynamics of ellipsoids show profound differences in pusher-type and puller-type active fluids. Figure~\ref{figure5}(a) shows the mean-squared displacements of ellipsoids along the major axis (MSD$_\text{a}$) and along the minor axis (MSD$_\text{b}$). Similar to the laboratory-frame MSDs, we found a superdiffusive regime at short times and a diffusive regime at long times in both MSD$_\text{a}$ and MSD$_\text{b}$. The diffusivity along the major and minor axis can be obtained by fitting the long-time diffusions with $\text{MSD}_\text{a} \equiv \langle \Delta\tilde{x}^2\rangle = 2D_at$ and $\text{MSD}_\text{b} \equiv \langle \Delta\tilde{y}^2\rangle = 2D_bt$, respectively. Both $D_a$ and $D_b$ increase with algal concentrations (Fig.~\ref{figure5}(b)). The bare diffusivity of Brownian ellipsoids in the body frame without algae is again orders of magnitude smaller, with $D_{a0} = 0.044$ $\mu$m$^2$/s and $D_{b0} = 0.033$ $\mu$m$^2$/s. 

\begin{figure*}
\begin{center}
\includegraphics[width=5.8in]{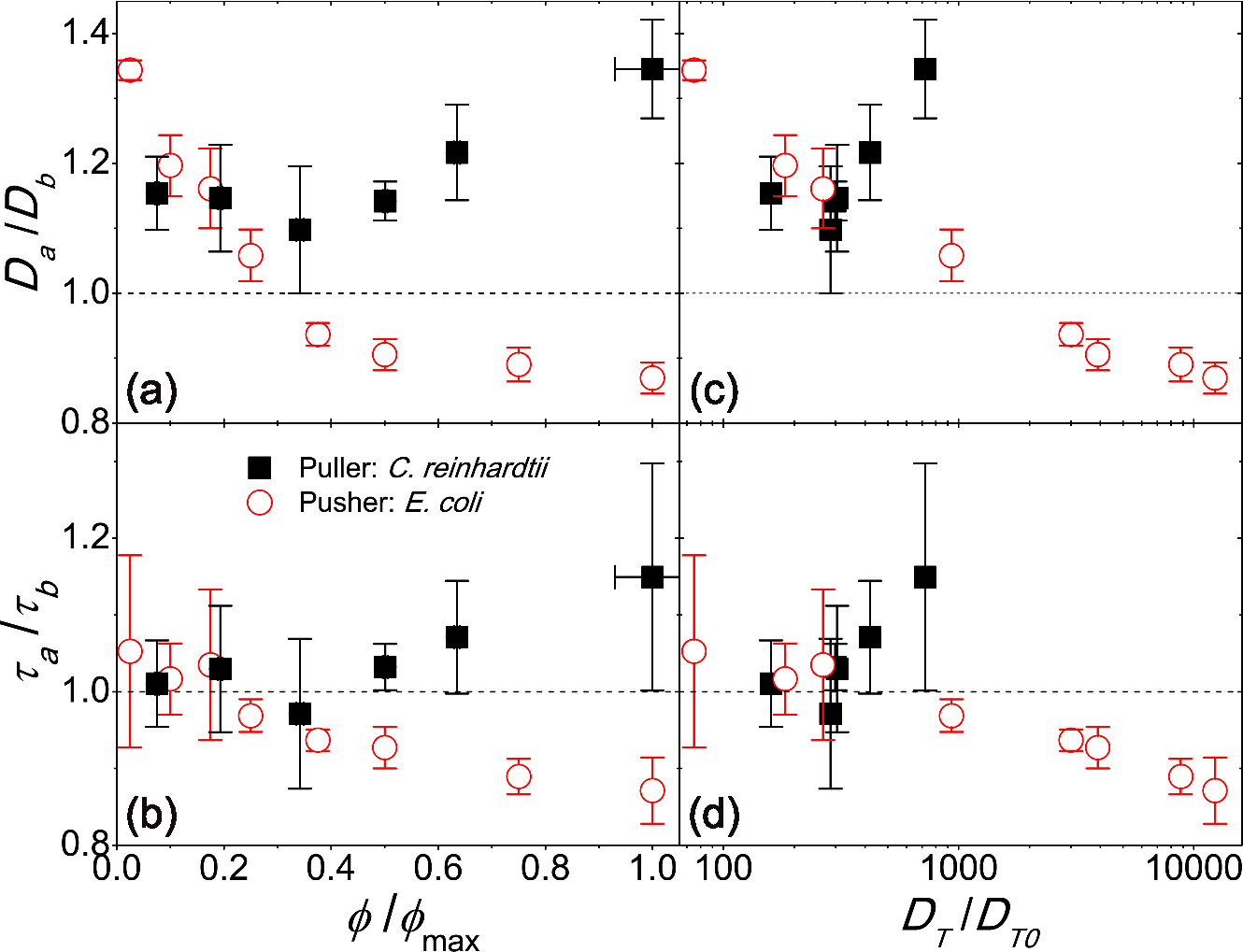}
\end{center}
\caption[Comparison of pushers and pullers]{Comparison of anisotropic diffusions in pullers and pushers. (a) Anisotropic diffusion of ellipsoids, $D_a/D_b$, in puller-type {\it C. reinhardtii}, and in pusher-type {\it E. coli}. The concentrations of active particles are normalized by the maximal concentration studied in experiments. $\phi_\text{max} = 3.9\%$ for {\it C. reinhardtii}. $\phi_\text{max} = 40 n_0$ for {\it E. coli}, where $n_0 = 8 \times 10^8$ cells/ml. (b) The ratio of the correlation times along the major and minor axis, $\tau_a/\tau_b$, in {\it C. reinhardtii} and in {\it E. coli}. Horizontal dashed lines indicate the ratio of 1. Since the maximal algal concentration we can achieve is much smaller than the maximal bacterial concentration due to different microbial physiology of {\it C. reinhardtii} and {\it E. coli} (see Sec IIB), it may seem arbitrary to normalize $\phi$ by $\phi_{max}$. Thus, we also plot $D_a/D_b$ (c) and $\tau_a/\tau_b$ (d) as a function of the normalized laboratory-frame translational diffusivity $D_T/D_{T,0}$, which show similar increasing and decreasing trends for pullers and pushers, respectively. The data for {\it E. coli} are extracted from Ref. \cite{Peng16}.} \label{figure6}
\end{figure*}

More interestingly, ellipsoids show an anisotropic diffusion with $D_a/D_b \neq 1$ (Fig.~\ref{figure5}(c)) \cite{footnote4}. The anisotropic diffusion in the body frame gives rise to a non-zero cross-correlation between the translation and rotation of particles in the laboratory frame \cite{Han06,Peng16}. Thus, the translational and rotational degrees of freedom of asymmetric tracers are coupled in algal suspensions. $D_a/D_b$ varies with algal concentrations. At low $\phi$, the diffusion along the major axis is faster than that along the minor axis with $D_a/D_b \approx 1.15$. As $\phi$ increases, the diffusion along the major axis is more strongly enhanced with $D_a/D_b$ increasing to $\sim 1.35$ at high $\phi$. As a comparison, for the same size Brownian ellipsoids, $D_{a0}/D_{b0}=1.33$. The upper limit of the anisotropic diffusion of Brownian ellipsoids in 3D is 2, which is reached when the aspect ratio of ellipsoids $p \to \infty$. More importantly, the increasing trend of $D_a/D_b$ with $\phi$ is opposite to the concentration dependence of the anisotropic diffusion of ellipsoids in pusher-type active fluids, where $D_a/D_b$ decreases monotonically with increasing $\phi$, to such an extent that $D_a/D_b$ becomes smaller than 1 at high enough concentrations (Figs.~\ref{figure6}(a)(c)) \cite{Peng16}. This difference manifests noticeably in the motion of ellipsoids: an ellipsoid diffuses fastest along the minor axis in pushers when $D_a/D_b < 1$, whereas it diffuses fastest along the major axis in pullers with $D_a/D_b > 1$. The origin of this striking difference will be discussed in Sec. III C below. As a final comment, it should be noted that since the dynamics of ellipsoids are dominated by the fluid flows of swimming algae, the contribution of the thermal fluctuation to the anisotropic diffusion is negligible in our experiments. Because $D_{a0}/D_{b0} = 1.33$ for Brownian ellipsoids, $D_a/D_b$ should eventually increase with diminishing algal contributions when the thermal fluctuation starts to play the dominant role. Hence, $D_{a}/D_{b}$ should show a non-monotonic trend as $\phi \to 0$. However, due to the small bare diffusivity of ellipsoids, any small drift currents in liquid films will induce large experimental uncertainties in our measurements. We were not able to resolve this increasing trend at small $\phi$ in our experiments.

\begin{figure}
\begin{center}
\includegraphics[width=3.2in]{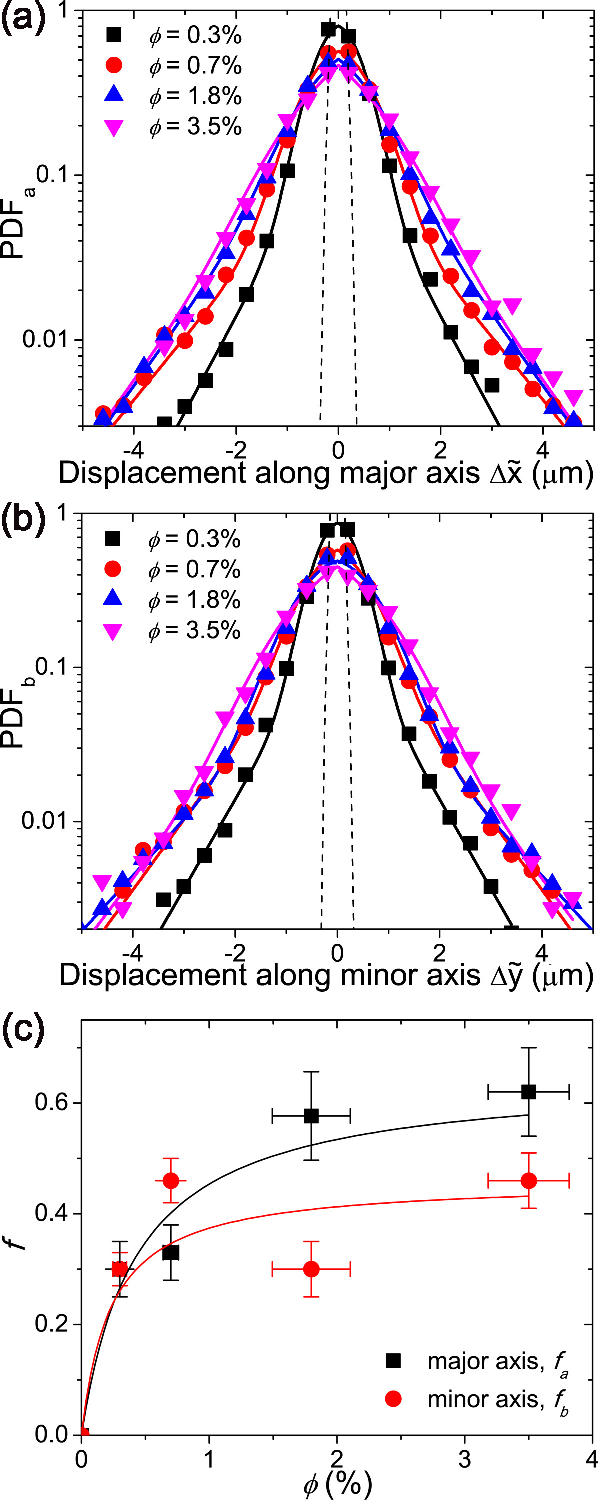}
\end{center}
\caption[Dynamics of ellipsoids in the body frame]{Probability distribution functions of body-frame translational displacements along the major axis (PDF$_\text{a}$) (a) and along the minor axis (PDF$_\text{b}$) (b) in a time interval $\Delta t = 0.1$ s at different algal concentrations $\phi$. The dashed lines are the PDFs of Brownian ellipsoids with bare diffusivity along the major and minor axis given in the text. The solid lines are fits using Eq.~\ref{equation1}. (c) The weighting factor, $f$, as a function of $\phi$ for both PDF$_\text{a}$ and PDF$_\text{b}$. The solid lines are visual guides.} \label{figure7}
\end{figure}

Last, for completeness, we also show the PDFs of ellipsoids in the body frame with a time interval $\Delta t = 0.1$ s. Figures~\ref{figure7}(a) and (b) show PDFs along the major axis (PDF$_\text{a}$) and along the minor axis (PDF$_\text{b}$), respectively. The results are qualitatively similar to PDF$_\text{T}$ in the laboratory frame. The body-frame PDFs can also be fitted with Eq.~\ref{equation1}. The resulting $f$ for PDF$_\text{a}$, $f_a$, and for PDF$_\text{b}$, $f_b$, are shown in Fig.~\ref{figure7}(c). $f_a$ and $f_b$ follow a qualitatively similar trend as $f_T$ and $f_R$, which increase with $\phi$ and saturate toward a constant at high $\phi$. $f_a$ is larger than $f_b$ at high $\phi$, indicating a stronger influence of advection on particles' motion along the major axis.    

\subsection{C. Discussions on the body-frame anisotropic diffusion}
We shall now discuss the origin of the anisotropic diffusion of ellipsoids in the body frame. From the Green-Kubo formula \cite{Zwanzig01}, the diffusivity of a random motion equals the integral of the velocity autocorrelation of the motion, 
\begin{equation}
D = \int_0^\infty dt \langle v(t_0+t) v(t_0) \rangle.
\label{equation3}
\end{equation}  
Assume the velocity autocorrelation follows a simple exponential decay, $\langle v(t_0+t) v(t_0) \rangle = \langle v_0^2 \rangle \exp(-t/\tau)$, where $\tau$ is the correlation time and $\langle v_0^2 \rangle$ is the mean-square velocity \cite{Peng16}. We have $D=\langle v_0^2 \rangle\tau$, a relation that can also be obtained based on dimensional analysis. $\langle v_0^2 \rangle^{1/2}$ indicates the step size of the random motion per unit time, whereas $\tau$ indicates the persistence of the motion. Naturally, a motion that has larger steps and is more persistent in its moving direction shows a large diffusivity. For Brownian ellipsoids, $\langle v_0^2 \rangle^{1/2} = \langle v_0 \rangle$ is the speed of Brownian particles in the ballistic regime \cite{Lukic05}. In this superdiffusive regime, we can further approximate $\langle v_0^2 \rangle \approx \langle \Delta x^2 \rangle/\Delta t^2$, where $\langle \Delta x^2 \rangle$ is the mean-square displacement in a small time interval $\Delta t$. 

\begin{figure}
\begin{center}
\includegraphics[width=3.2in]{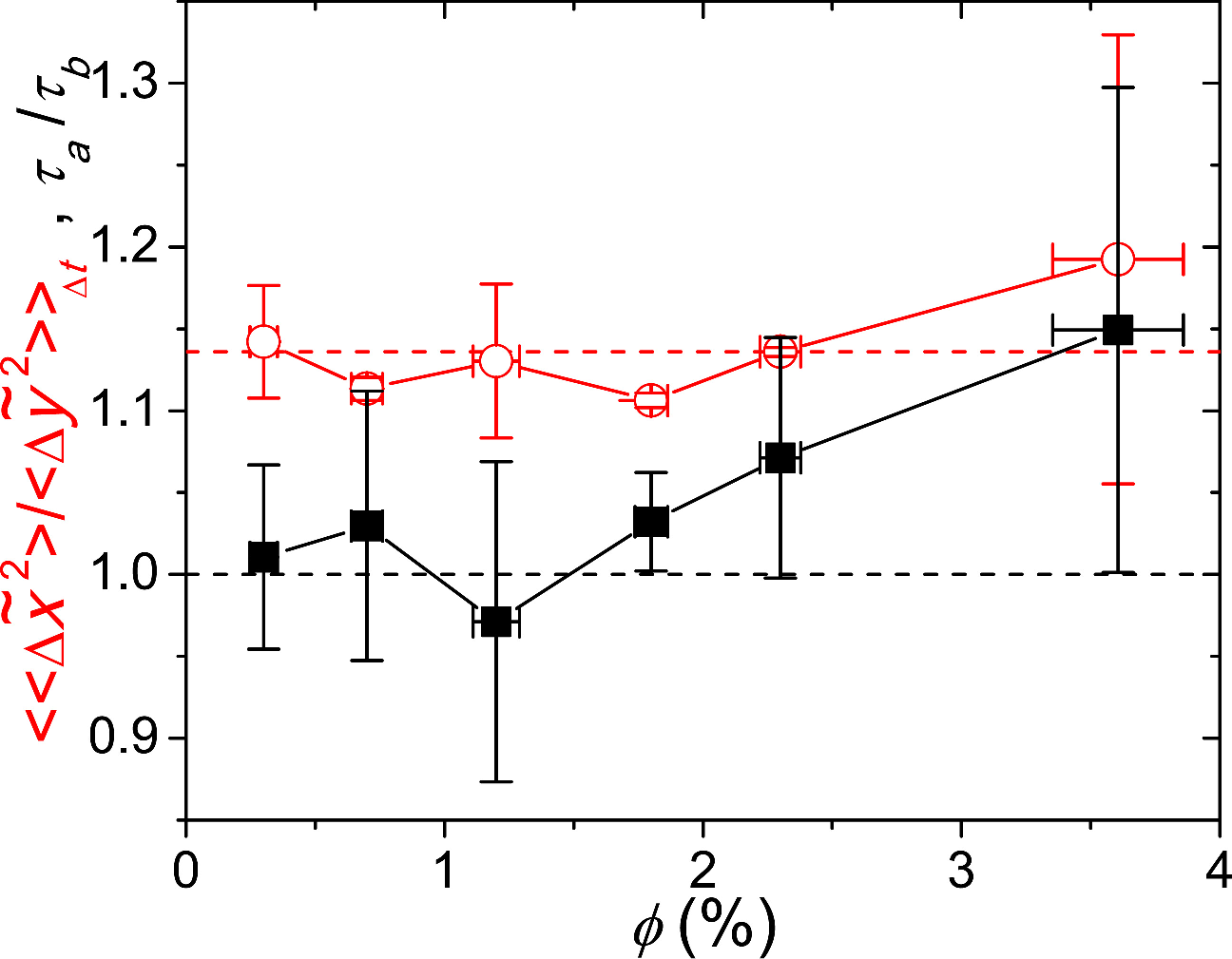}
\end{center}
\caption[Origin of anisotropic diffusion]{Origin of the anisotropic diffusion. Black squares are the ratio of the correlation times along the major and minor axes, $\tau_a/\tau_b$, as a function of algal concentrations $\phi$. Red circles are the average ratio of the mean-square displacements along the major and minor axis in the superdiffusive regime, $\langle \langle \Delta \tilde{x}^2 \rangle/\langle \Delta \tilde{y}^2 \rangle \rangle_{\Delta t}$, as a function of $\phi$. The horizontal dashed lines indicate  $\langle \langle \Delta \tilde{x}^2 \rangle/\langle \Delta \tilde{y}^2 \rangle \rangle_{\Delta t}$ = 1.13 (upper, red) and $\tau_a/\tau_b = 1$ (lower, black), respectively.} \label{figure8}
\end{figure}

Applying the above general consideration in the anisotropic diffusion of ellipsoids in active fluids, we have 
\begin{equation}
\frac{D_a}{D_b} = \frac{\langle \Delta \tilde{x}^2 \rangle}{\langle \Delta \tilde{y}^2 \rangle}\cdot\frac{\tau_a}{\tau_b},
\label{equation4}
\end{equation}     
which leads to 
\begin{equation}
\frac{\tau_a}{\tau_b} = \frac{D_a/D_b}{\langle \Delta \tilde{x}^2 \rangle/\langle \Delta \tilde{y}^2 \rangle}.
\label{equation5}
\end{equation}
Here, $\tau_a$ and $\tau_b$ are the correlation times of ellipsoids' motions along the major and minor axis, respectively. We measured the average ratio of the mean-square displacements in the superdiffusive regime, $\langle \langle \Delta \tilde{x}^2 \rangle/\langle \Delta \tilde{y}^2 \rangle \rangle_{\Delta t}$, at different $\phi$, where the average is taken for all $\Delta t \leq  0.8$ s, the upper limit of the superdiffusive regime (Fig.~\ref{figure5}(a)). We found that $\langle \langle \Delta \tilde{x}^2 \rangle/\langle \Delta \tilde{y}^2 \rangle \rangle_{\Delta t} \approx 1.13 \pm 0.02$ independent of $\phi$ at low $\phi$ and increases slightly at the highest $\phi$ (Fig.~\ref{figure8}). With $\langle \langle \Delta \tilde{x}^2 \rangle/\langle \Delta \tilde{y}^2 \rangle \rangle_{\Delta t}$ and $D_a/D_b$, we obtained $\tau_a/\tau_b$ from Eq.~\ref{equation5} (Fig.~\ref{figure8}). Although $\tau_a/\tau_b$ thus obtained shows large errors due to drift flows in the thin film and the variation of film thickness and algal activity in different experimental runs, the increasing trend of the mean value of $\tau_a/\tau_b$ is clear. 

The decomposition of $D_a/D_b$ into $\langle \langle \Delta \tilde{x}^2 \rangle/\langle \Delta \tilde{y}^2 \rangle \rangle_{\Delta t}$ and $\tau_a/\tau_b$ in Eq.~\ref{equation4} helps to illustrate the origin of the anisotropic diffusion of ellipsoids. First, the step size along the major axis is larger than that along the minor axis. However, the ratio between the two, $\langle \langle \Delta \tilde{x}^2 \rangle/\langle \Delta \tilde{y}^2 \rangle \rangle_{\Delta t}$, keeps roughly constant except at the highest $\phi$. The result is consistent with the previous observation of $f_a/f_b$, where advection exerts a stronger influence on the motion of particles along the major axis at small time intervals (Fig.~\ref{figure7}(c)). The ratio of the correlation times, $\tau_a/\tau_b$, shows a more interesting trend. At low $\phi$, $\tau_a \approx \tau_b$, indicating a similar persistence for the motions along the major and minor axes. However, $\tau_a/\tau_b$ increases with $\phi$ at high $\phi$, which leads to the increase of $D_a/D_b$ with $\phi$. Hence, the anisotropic diffusion of ellipsoids in algal suspensions arises from the increase of the persistence of the motion along the major axis relative to that along the minor axis. This observation is in sharp contrast with the dynamics of ellipsoids in {\it E. coli} suspensions, where $\tau_a/\tau_b$ decreases monotonically with bacterial concentrations (Figs.~\ref{figure6}(b)(d)) \cite{Peng16}. As a comparison, the correlation time of thermal Brownian motions is given by the inertial time of Brownian particles, $\tau = m/\zeta$, where $m$ is the mass of the particles and $\zeta$ is the drag coefficient \cite{Zwanzig01}. Hence, $\tau_{a0}/\tau_{b0} = \zeta_b/\zeta_a = D_{a0}/D_{b0} = 1.33$. $\zeta_{a,b}$ are the drag coefficients along the major and minor axes of ellipsoids, which are related to the anisotropic diffusion through the Stokes-Einstein relation. $\tau$ is on the order of micro-seconds beyond the time resolution of our experiments \cite{footnote5}. We should again emphasize that the origins of the correlation time of Brownian diffusion and enhanced diffusion are completely different. Thus, the limit of zero algal concentrations $(\phi \to 0)$ is not equivalent to Brownian diffusion with $\phi = 0$.       

Therefore, to understand the effect of pushers and pullers on the dynamics of asymmetric tracers, it is important to reveal how the swimming of microswimmers affects the correlation times of ellipsoids' motions. The simple hydrodynamic calculation by Peng {\it et al.} provides a qualitative guideline \cite{Peng16}. First, the motion of an ellipsoid in the dipole flow of a single microswimmer can be calculated in Stokes flow as \cite{Kim91}  
\begin{equation}
\mathbf{v_p} = \mathbf{u},
\label{equation6}
\end{equation}  
\begin{equation}
\mathbf{\omega_p} = \frac{1}{2}\nabla \times \mathbf{u}+\frac{p^2-1}{p^2+1}\hat{a} \times (\mathbf{\epsilon}\cdot\hat{a}).
\label{equation7}
\end{equation}  
Here, $\mathbf{v_p}$ and $\mathbf{\omega_p}$ are the translational and angular velocity of the ellipsoid with aspect ratio $p$. The ellipsoid is treated as a force-free and torque-free point particle. The assumption is valid in the far field when the concentration of microswimmers is low. $\mathbf{u} = -k (\mathbf{r}/r^3-3x^2\mathbf{r}/r^5)$ is the dipole flow at $\mathbf{r} = (x,y)$ induced by a microswimmer at origin. The dipole strength, $k$, is positive for pushers and negative for pullers. $\mathbf{\epsilon}$ is the rate-of-strain tensor of the dipole flow. $\hat{a}$ is the unit vector along the major axis of the ellipsoid. In the dilute limit, the motion of the ellipsoid under the influence of many microswimmers can be obtained by averaging all the possible orientations and positions of microswimmers relative to the ellipsoid---a mean-field approximation that ignores the correlation between microswimmers. The method has been successfully used for interpreting the linear relation between the enhanced diffusivity of spherical tracers and the concentration of active particles \cite{Lin11,Morozov14}. A coupling scaler, $S \equiv \mathbf{\omega_p}\cdot \left(\frac{\hat{a}'\times\mathbf{v_p}}{|\hat{a}'\times\mathbf{v_p}|} \right)$ can then be defined, which quantifies the intrinsic coupling between the translation and rotation of an ellipsoid in the dipole flow \cite{Peng16}. $\hat{a}'$ is a unit vector that satisfies $\hat{a}'= \hat{a}$ when $\hat{a}'\cdot\mathbf{v}\geq 0$ and $\hat{a}'= - \hat{a}$ when $\hat{a}\cdot\mathbf{v}<0$. In other words, $\hat{a}'$ is along the major axis that always forms an acute angle with $\mathbf{v}$. From the definition, the amplitude of $S$ characterizes the speed of the particle rotation, $|S|=|\mathbf{\omega_p}|$, and the sign of $S$ indicates the direction of the rotation. A negative $S$ corresponds to a rotation that tends to align the minor axis of the ellipsoid along with its translational direction $\mathbf{v_p}/|\mathbf{v_p}|$, whereas a positive $S$ corresponds to a rotation that aligns the major axis of the ellipsoid along with $\mathbf{v_p}/|\mathbf{v_p}|$. By taking an average over all possible orientations and positions of microswimmers, Peng {\it et al.} showed that the average of $S$ follows
\begin{equation}
\langle S \rangle = -\frac{p^2-1}{p^2+1}\frac{3k}{V_0}\phi,
\label{equation8}
\end{equation}
where $V_0$ indicates the volume of microswimmers. $\langle S \rangle = 0$ for spherical tracers with $p = 1$. More importantly, $\langle S \rangle > 0 $ for pullers and $\langle S \rangle < 0 $ for pushers. 

The effect of a non-zero $\langle S \rangle$ is equivalent to a straining flow, which rotates an asymmetric tracer in a direction depending on the swimming mechanism of microswimmers and the orientation of the tracer with respect to its translational direction. The dynamics of asymmetric tracers can thus be modeled as the diffusion of the tracers in a thermal bath with an effective temperature $T_\text{eff}$ under the influence of an external rotational potential characterized by $\langle S \rangle$ \cite{Peng16}. The correlation time of the diffusion can be obtained by solving an over-damped Langevin equation with a rotational potential, which leads to $\tau_a = t_c/(1-2t_c\langle S \rangle/\pi)$ and $\tau_b = t_c/(1+ 2t_c\langle S \rangle/\pi)$. Here, $t_c$ indicates the correlation time of the random fluctuation of the effective thermal bath, which is related to the transition time between the superdiffusive regime to the diffusion regime (Fig.~\ref{figure5}(a)) \cite{Wu00}. Thus, we finally have
\begin{equation}
\frac{\tau_a}{\tau_b} = \frac{1+\text{sgn}\left( \langle S \rangle \right) t_c/t_0}{1-\text{sgn}\left(\langle S \rangle\right)t_c/t_0},
\label{equation9}
\end{equation} 
where $t_0 \equiv \frac{\pi/2}{|\langle S \rangle|}$ is a characteristic time for the rotation of tracers over an angle of $\pi/2$ due to a non-zero $|\langle S \rangle| = |\mathbf{\omega_p}|$. Note that $\text{sgn}(x) = 1$ when $x > 0$, $\text{sgn}(x) = -1$ when $x < 0$ and $\text{sgn}(x) = 0$ when $x = 0$. From Eq.~\ref{equation8}, $t_0 \sim 1/|\langle S \rangle|$ should decrease with increasing $\phi$. Thus, for pushers, since $\langle S \rangle < 0$, $\tau_a/\tau_b = (1-t_c/t_0)/(1+t_c/t_0) \leq 1$ and decreases with increasing $\phi$, a prediction that has been confirmed experimentally from the study of the anisotropic diffusion of ellipsoids in {\it E. coli} suspensions (Figs.~\ref{figure6}(b)(c)) \cite{Peng16}. For pullers, $\langle S \rangle > 0$. Thus, $\tau_a/\tau_b = (1+t_c/t_0)/(1-t_c/t_0)\geq 1$, which should increase with increasing $\phi$. Our experiments with {\it C. reinhardtii} suspensions qualitatively agree with this prediction (Fig.~\ref{figure8}). Finally, for spherical tracers, $\langle S \rangle = 0$. $\tau_a = \tau_b$, independent of whether active fluids are pushers or pullers.

\section{IV. Conclusions}
We have studied the dynamics of ellipsoids immersed in {\it C. reinhardtii} suspensions, a premier model of puller-type active fluids. Different from spherical tracers that have been extensively studied, ellipsoidal tracers possess an additional rotational degree of freedom and, therefore, show much richer dynamics. 

In the laboratory frame, both translation and rotation of ellipsoids show superdiffusive motions at short times and enhanced diffusions at long times, similar to the dynamics of spherical tracers. By analyzing the probability distribution functions of the displacements of ellipsoids, we showed that the translational and rotational motions of ellipsoids can be quantitatively understood from the balance of advection and diffusion induced by the swimming of algae. Due to the nature of the dipole flow of microswimmers, the translational advection shows a longer range of influence than the rotational advection. 

The body-frame dynamics of ellipsoids reveal the distinct difference between pushers and pullers. Although when viewed independently the motions of ellipsoids along the major and minor axes show qualitatively similar trends, ellipsoids in the body frame show an unusual anisotropic diffusion, where the long-time diffusion along the major axis is more strongly enhanced with increasing algal concentrations than that along the minor axis. Such an anisotropic diffusion dictates a coupling between the translation and rotation of ellipsoids in the laboratory frame. By decomposing diffusivity into different components, we demonstrated that the anisotropic diffusion arises from the differential variation of the correlation times of ellipsoids' motions along the major and minor axes. The motion of an ellipsoid along its major axis becomes more persistent when algal concentration increases. This trend is in sharp contrast to the dynamics of ellipsoids in pusher-type {\it E. coli} suspensions, where the persistence of the motion along the major axis decreases with increasing bacterial concentrations. This sharp difference can be qualitatively explained at the mean-field level by considering the translation-rotation coupling induced by the straining component of the dipole flow of microswimmers. As such, our study showed that the anisotropic diffusion of asymmetric tracers is a universal feature of active fluids, which can be used as a convenient experimental indicator for distinguishing pushers versus pullers.

Finally, our work also provided experimental results on the dynamics of asymmetric particles in suspensions of eukaryotic microorganisms. Since asymmetric particles (e.g. macromolecules and dead bodies of microorganisms) are naturally more abundant than spherical particles, our results should be more relevant to realistic biological systems and be useful for understanding the physiology of swimming eukaryotes.

\section{Acknowledgments}
We acknowledge P. Lefebvre for providing us with the $C. reinhardtii$ strain and helping us with algae culturing. We also thank B. Zhang, L. Gordillo and D. Samanta for help with experiments and L. Lai for fruitful discussions on theory. The research was supported by ACS Petroleum Research Fund (54168-DNI9) and by the David $\&$ Lucile Packard Foundation. C. T. thanks support from the Coating Process Fundamentals Program (CPFP) at University of Minnesota. X. X. acknowledges support by the National Natural Science Foundation of China No. 11575020.


\begin{thebibliography}{99}
\bibitem{footnote4} Note that when evaluating the experimental errors of $D_a/D_b$, we computed $D_a/D_b$ for each experimental run and then calculated the standard deviation of the ratios of all runs. Since the variations of $D_a$ and $D_b$ are correlated, the errors of $D_a/D_b$ are significantly smaller than that evaluated by treating $D_a$ and $D_b$ as independent variables.
\bibitem{footnote5} At such a small time scale, fluid inertial certainly plays an important role \cite{Lukic05}. It is not clear how the fluid inertial differentially affects the motion of ellipsoids along the major and minor axis and modifies $\tau_{a0}/\tau_{b0}$.
\end{thebibliography}
\end{document}